\documentclass[preprint]{revtex4-1}
\usepackage{color}
\usepackage{amstext}
\usepackage{amssymb}
\usepackage{graphicx}
\usepackage[normalem]{ulem}

\begin{document}

\title{Error field penetration and locking to the backward propagating wave}

\author{John~M.~Finn$^{1}$, Andrew~J.~Cole$^{2}$ and Dylan~P.~Brennan$^{3}$}
\address{$^1$ Theoretical Division, Los Alamos National Laboratory, Los Alamos, NM \\ $^2$ Department of Applied Mathematics and Applied Physics, Columbia University, New York, NY \\ $^3$ Plasma Physics Laboratory, Princeton University, Princeton, NJ}

\begin{abstract}
Resonant field amplification, which may lead to locking or error field penetration, involves driving
a weakly stable tearing perturbation in a rotating toroidal plasma.
In this paper it is shown that the locking characteristics for modes
with finite real frequencies $\omega_{r}$ are quite different from
the conventional results. A calculation of the tearing mode amplitude
assuming modes with frequencies $\pm\omega_{r}$ in the plasma frame
shows that it is maximized when the frequency of the stable backward
propagating mode ($-\omega_{r}$) is zero, i.e.~when
$v=\omega_{r}/k$. Even more importantly, the locking torque is \emph{exactly}
zero at the mode phase velocity, with a pronounced peak at just higher
rotation, leading to a locked state with plasma velocity $v$ just
above the mode phase velocity in the lab frame. Real frequencies are known to occur due to the Glasser effect\cite{GGJ1,GGJ2} for modes in the resistive-inertial
(RI) regime. This therefore leads to locking of the plasma velocity
to just above the phase velocity. It is also shown here that real
frequencies occur over a wide range of parameters in the visco-resistive
(VR) regime with pressure, and the locking torque is similar to that
in the RI regime. The same is true in other tearing regimes, with real frequencies due to diamagnetic propagation. Other nonlinear effects and the possibility of applying external fields of different helicities
to drive sheared flows in toroidal plasmas are discussed.
\end{abstract}

\maketitle

\section{Introduction}

Resonant field amplification\cite{BoozerRFA} (RFA) is the response
of a magnetically confined toroidal plasma to an externally applied
nonaxisymmetric magnetic field, an \emph{error field}.  Error fields are generally unavoidable, and in some cases such fields can be applied purposefully, as discussed below.  RFA can lead to error field penetration (EFP) (EFP)\cite{Fitzpatrick1,Fitzpatrick2,FitzpatrickEtAl93,FitzpatrickHender,
Gimblett,NaveWesson} or locking. The accepted view is that EFP involves locking the
plasma rotation to just above zero velocity in response to the error field. If this occurs it can cause a loss of the benefit of sheared rotation on plasma stability, and
on occasions can lead to disruptions in tokamaks. It is well known
that RFA occurs by means of coupling to a stable
tearing mode, and that the response is maximized when the tearing
mode is near marginal stability\cite{Fitzpatrick1,Fitzpatrick2,BoozerRFA,FitzpatrickEtAl93,
Gimblett,NaveWesson}.
The dispersion relation for a spontaneous mode (with $\tilde{\psi}(r_{w})=0$)
is $\Delta(\gamma_{d})=\Delta'$, where $\gamma_{d}$ is the growth
rate $\gamma+ikv$ in the plasma frame and $\Delta'$ and $\Delta$ are the usual constant-$\psi$ matching
parameters from the tearing mode outer and inner regions, respectively. Here, $v$ is the toroidal plasma velocity ($E\times B$ plus parallel velocity) at the mode rational surface. The steady-state magnetic perturbation at the edge of the tearing layer at the mode rational surface
($k_{||}=0$) is called the \emph{reconnected flux} in the context of the constant-$\psi$ approximation, and is known to have the form
\begin{equation}
\tilde{\psi}(r_{t})=-\frac{l_{21}}{\Delta'-\Delta(ikv)}\tilde{\psi}(r_{w}).\label{eq:psit-from-psiw}
\end{equation}
Here, $\tilde{\psi}(r_{w})$ is the amplitude of the error field applied
at the wall, $r_{t}$ is the radius of the mode rational surface,
  and $l_{21}$ is an inductance
coefficient. We describe the results here in terms of large aspect ratio cylindrical geometry,
although one toroidal effect, namely favorable average curvature for safety factor $q>1$, is of importance and is included. The
quantities $\Delta$, $\Delta'$ and $l_{21}$ are normalized so that
the plasma radius is equal to unity. 

For tearing modes in the viscoresistive (VR) regime\cite{CGJ,ColeFitzpatrick2},
with resistivity $\eta$ and ion viscosity $\mu$, $\Delta(\gamma)$
has the form 
\begin{equation}
\Delta(\gamma)=\frac{\mu^{1/6}}{\eta^{5/6}|k_{||}'|^{1/3}B^{1/3}}\Delta_{s}\gamma=\gamma\tau_{vr},\label{eq:Delta-gamma-for-VR}
\end{equation}
where $\Delta_{s}$ is a dimensionless positive constant of order
unity. From Eq.~(\ref{eq:Delta-gamma-for-VR})
it is seen that spontaneous VR tearing modes have real $\gamma$ for
all $\Delta'$ and Eq.~(\ref{eq:psit-from-psiw}) shows that the largest response to $\tilde{\psi}(r_{w})$
for stable plasmas is near marginal stability, where $|\Delta'-ikv\tau_{vr}|$
is smallest. This occurs for $\Delta'\rightarrow0-$ with $-\omega_r+kv=0$, i.e.~$v=0$. 

\begin{figure}
\includegraphics[scale=0.4]{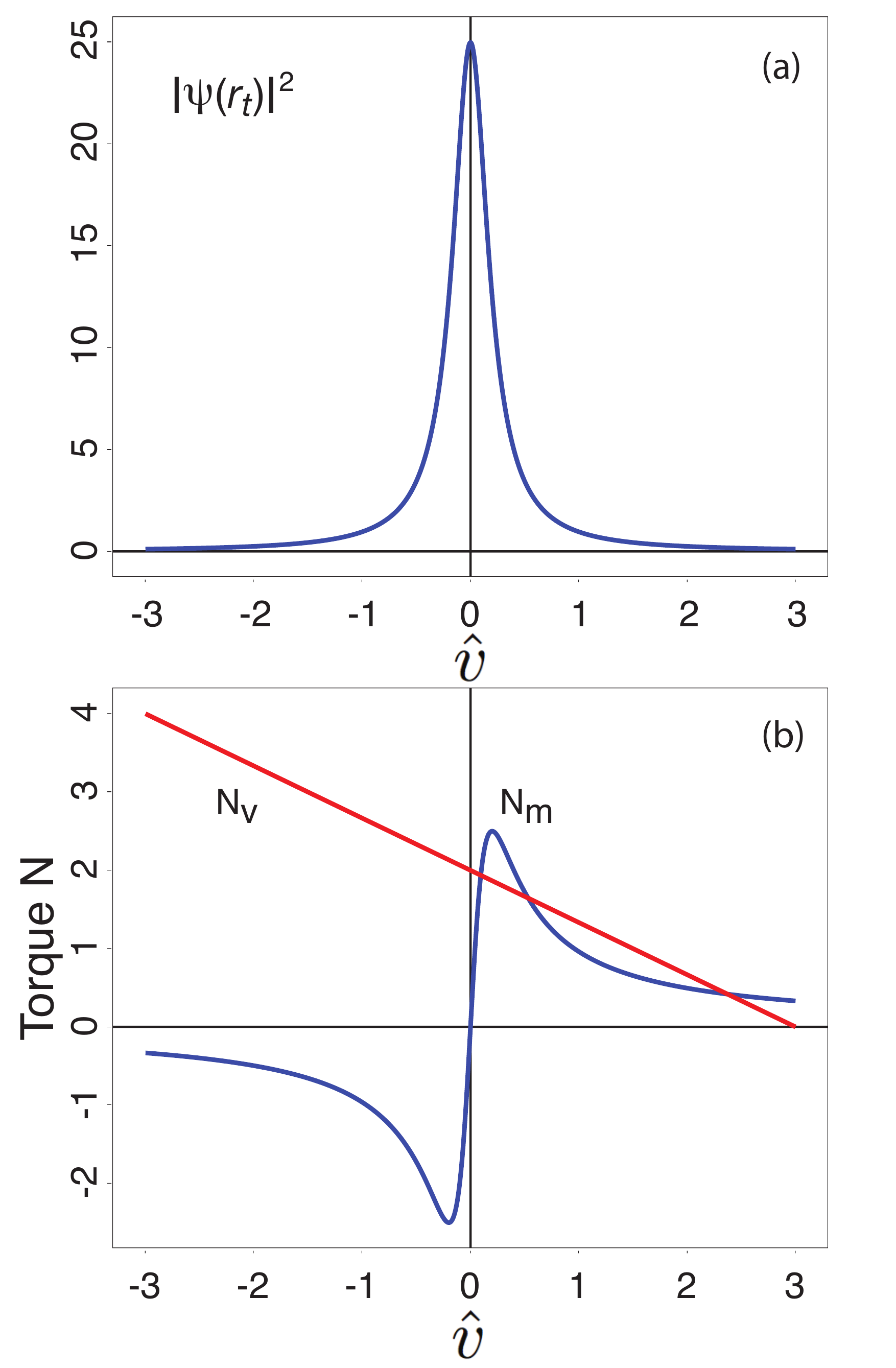}
\caption{Plot (a) of reconnected flux magnitude $|\tilde{\psi}(r_{t})|^{2}$
vs.~$\hat{v}\equiv kv\tau_{vr}$ for a weakly stable VR mode.
In (b) are shown the torque $-N_{m}$ vs.~$\hat{v}$ as well
as the viscous torque curve $N_{v}\propto v_{0}-v$. In this parameter
range there are three equilibria, with the locked equilibrium having
very small $\hat{v}\gtrsim0$.}
\label{fig:TorqueCurveVR}
\end{figure}

The toroidal torque on the plasma, which comes from the Maxwell stress
in the non-ideal layer, takes the quasilinear form\cite{Fitzpatrick1,Fitzpatrick2}
\begin{equation}
N_{m}=-\frac{k}{2}|\tilde{\psi}(r_{t})|^{2}\text{Im}\Delta(ikv).\label{eq:Torque-Im-Delta}
\end{equation}
(This quantity is the integral over the tearing layer of $\tilde{j}_{||} \tilde{B}_r$ and is formally a force in this reduced model.  In a tokamak, because of strong damping of the poloidal flow, the toroidal force component is most important, which gets multiplied by the factor $r_t/q(r_t)R$, where $q$ is the safety factor and $R$ is the toroidal major radius.  Thus there is a direct relation between $N_m$ and torque along the tokamak symmetry axis from toroidal force.) Using Eq.~(\ref{eq:psit-from-psiw}) it is readily seen that this torque in the VR regime is
\begin{equation}
N_{m}=-\frac{k^{2}l_{21}^{2}|\tilde{\psi}(r_{w})|^{2}}{2}\frac{v\tau_{vr}}{\Delta'^{2}+k^{2}v^{2}\tau_{vr}^{2}},\label{eq:Torque-conventional-Lorentzian}
\end{equation}
behaving as $-v/(v_{1}^{2}+v^{2})$, with extrema at $k^{2}v^{2}=\Delta'^{2}/\tau_{vr}^{2}$,
and symmetry about $v=0$, $N_{m}(-v)=-N_{m}(v)$.  The quantities $|\tilde{\psi}(r_{t})|^{2}$ and $N_{m}$ as a function of $\hat{v}\equiv kv\tau_{vr}$ are shown in Fig. \ref{fig:TorqueCurveVR}, illustrating the behavior around $v=0$.

\begin{figure}
\includegraphics[scale=0.6]{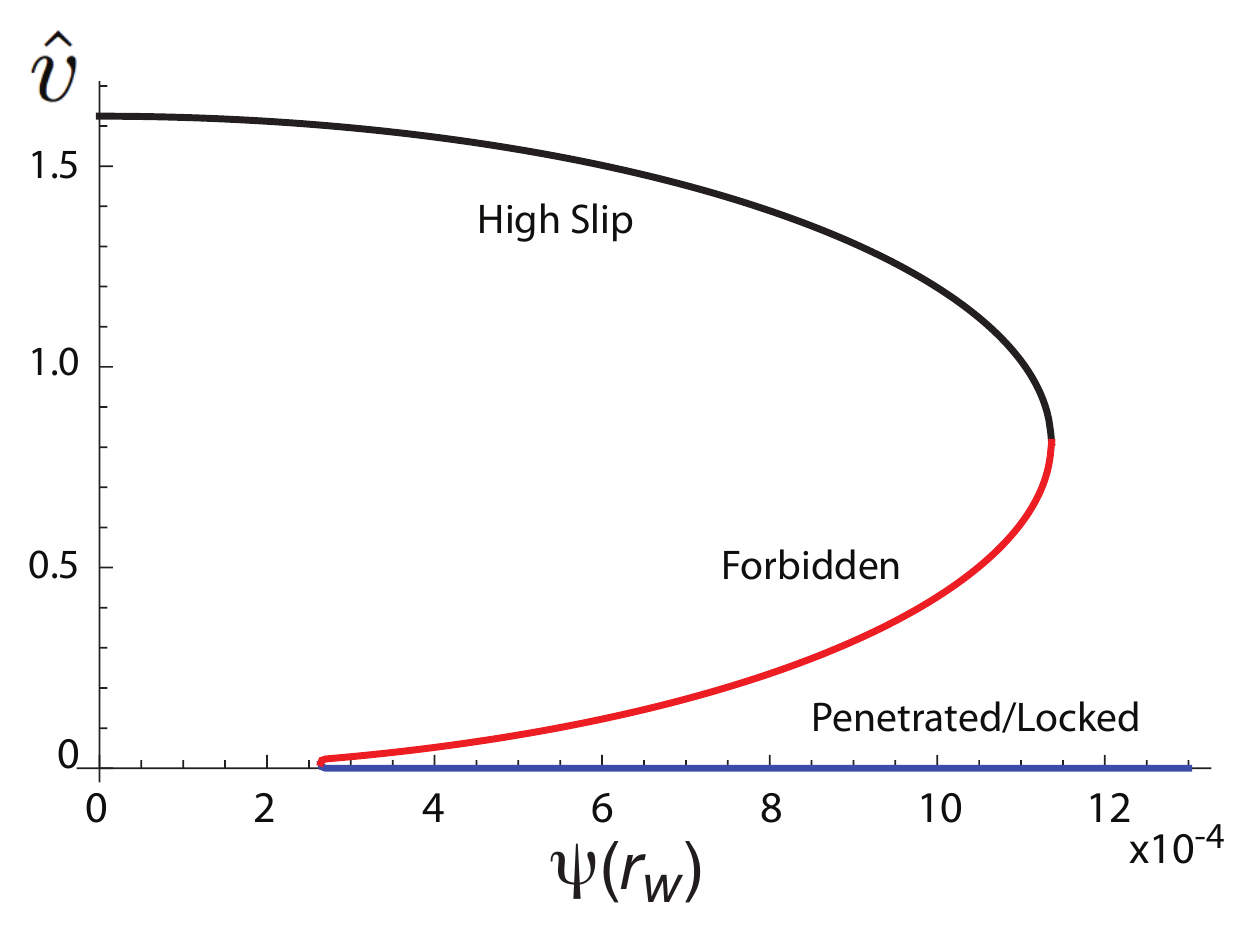}
\caption{Bifurcation diagram for the VR regime, a plot of the the equilibrium flow vs.~the wall perturbation amplitude.}
\label{fig:bifVR}
\end{figure}

Based on these results, it has been shown\cite{Fitzpatrick1,Fitzpatrick2,BoozerRFA,FitzpatrickEtAl93}
that models with a viscous response torque at the layer $N_{v}=N_{0}(v_{0}-v)$ (where $N_{0}$ represents a balance between the momentum source and viscous drag, with no error field to give an equilibrium flow $v_{0}$) can have
three rotation states, intersections with $N_{m}+N_{d}=0$, for intermediate
values of $|\tilde{\psi}(r_{w})|$, as shown in Fig. \ref{fig:TorqueCurveVR}b. The bifurcation diagram illustrating these states is shown in Fig.~\ref{fig:bifVR}. The fastest is the rapidly rotating or high slip
state, and the slowest is the penetrated or \emph{locked state}. The middle intersection
represents an unstable or forbidden rotation. For small $|\tilde{\psi}(r_{w})|$
or large $N_{0}$ (or large $v_0$), only the rapidly rotating state exists; for increasing
$|\tilde{\psi}(r_{w})|$ or decreasing $N_{0}$ there is a bifurcation
at $\tilde{\psi}_{w1}$ past which the locked and intermediate states
can exist. For further increasing $|\tilde{\psi}_{w}(r_{w})|$ or
decreasing $N_{0}$, there is a bifurcation at $\tilde{\psi}_{w2}$
past which the two upper states coalesce and only the locked state
can exist. Since $\tilde{\psi}_{w2}>\tilde{\psi}_{w1}$, these bifurcations
have hysteresis\cite{Fitzpatrick1,Fitzpatrick2,FinnSovinec,FinnLocking,Gimblett}.

\section{Resistive-inertial (RI) regime}
\label{sec:ri}
In the resistive-inertial (RI) regime, the quantity $\Delta$ takes
the form 
\begin{equation}
\Delta(\gamma)=\frac{\rho^{1/4}}{\eta^{3/4}|k_{||}'|^{1/2}B^{1/2}}\Delta_{s}\gamma^{5/4}=\gamma^{5/4}\tau_{ri}^{5/4},\label{eq:RI-Delta-of-gamma}
\end{equation}
where $\Delta_{s}$ is another positive dimensionless constant. In
the presence of equilibrium pressure gradient $p'(r_{t})$ and field
line curvature at the tearing layer, parallel dynamics, perpendicular
compression and particle transport, it is known that for sufficiently
high plasma $\beta$ this is modified to give\cite{CGJ,FinnManheimer}
\begin{equation}
\Delta_{s}\approx\left(A-B/\gamma^{3/2}\right),\label{eq:Deltas-for-RI-D}
\end{equation}
where $A$ and $B\propto D_{s}$ are constant and $D_{s}=-2rp'/B_{\theta}^{2}R^{2}q'^{2}$
is the Suydam parameter. For $D_{s}>0$ ($B>0$),
this result leads to the observation that tearing modes are destabilized
for all values of $\Delta'$ \cite{CGJ,FinnManheimer}, becoming electrostatic
with $\gamma\sim(D_{s}/\Delta')^{4}$ as $\Delta'\rightarrow-\infty$ \cite{FinnManheimer}.
In toroidal geometry $D_{s}$ is replaced by the Mercier parameter
$D=-2rp'(1-q^{2})/B_{\theta}^{2}R^{2}q'^{2}=(1-q^{2})D_{s}$, giving
$B<0$. Again, $q$ is the safety factor and $R$
is the toroidal major radius. The $1-q^{2}$ factor is due to the
average curvature in toroidal geometry and leads to $D<0$ for $q(r_{t})>1$
and $p'(r_{t})<0$.

For favorable curvature $D<0$ the dispersion relation $\Delta(\gamma)=\Delta'$
gives two real roots with $\gamma>0$ for $\Delta'>\Delta_{min}\propto|D|^{5/6}$.
For $\Delta'<\Delta_{min}$, there are complex conjugate roots with
$\gamma=\gamma_{r}\mp i\omega_{r}$ in the plasma frame; these roots
have $\gamma_{r}<0$ (are stable) if $\Delta'<\Delta_{c}$; the quantity
$\Delta_{c}$ is also proportional to $|D|^{5/6}$.\cite{GGJ1,GGJ2}\\

\begin{figure}
\includegraphics[scale=0.35]{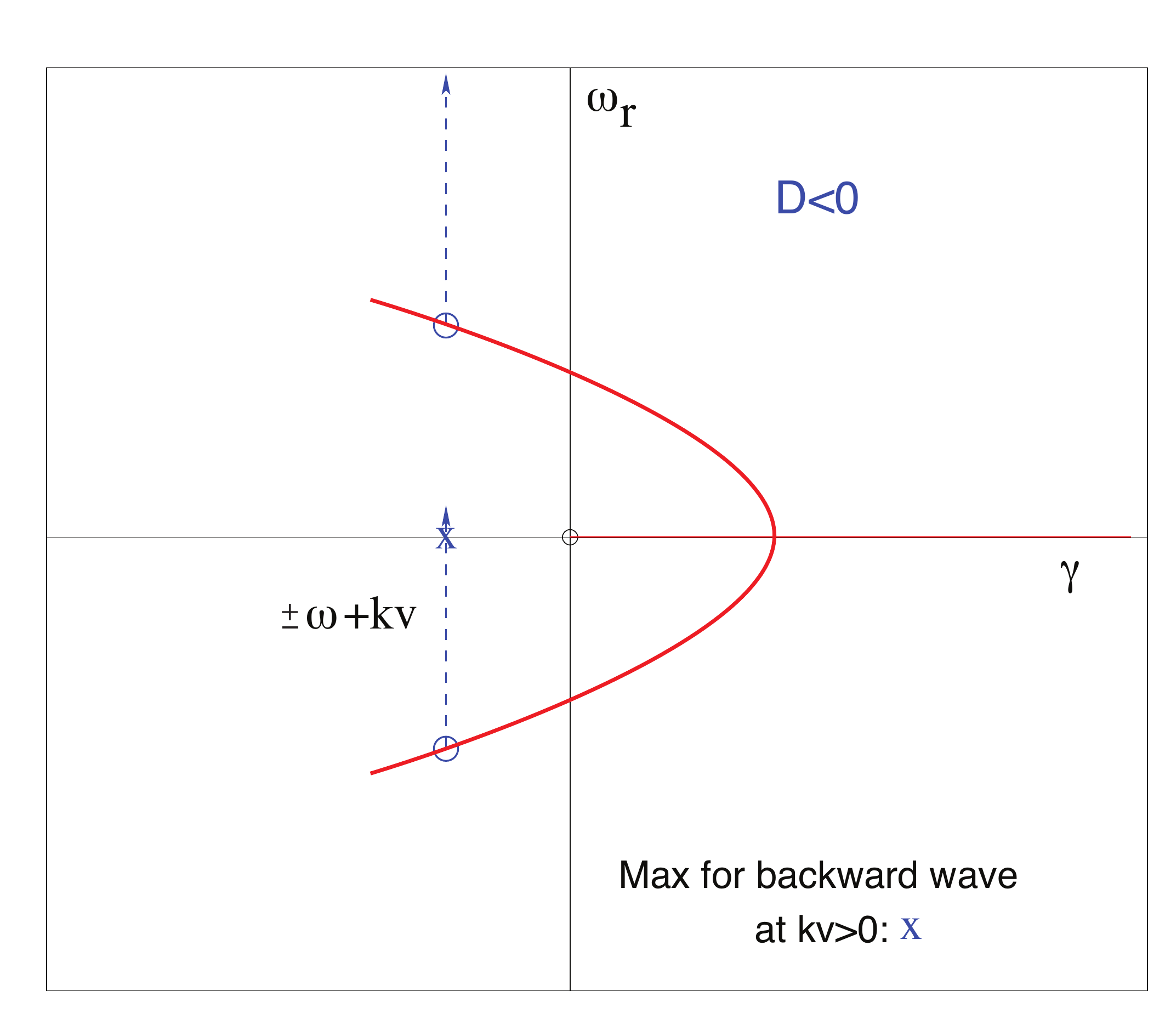}
\caption{Sketch of the locus of roots in the RI regime with favorable curvature. The blue arrows indicate the change in the roots as $\hat{v}$ increases, with the backward wave complex frequency coming closest to zero for $v=\omega_r/k$.}
\label{fig:Locus-of-roots-RI}
\end{figure}

The possibility of complex conjugate roots (roots for the spontaneous
modes with real frequency) near marginal stability $\gamma_{r}\lesssim0$
or $\Delta'\lesssim\Delta_{c}$, leads to the main result in this
paper. The reconnected flux from Eq.~(\ref{eq:psit-from-psiw}) satisfies
\[
|\tilde{\psi}(r_{t})|^{2}=\frac{l_{21}^{2}}{|\Delta'-\Delta(ikv)|^{2}}|\tilde{\psi}(r_{w})|^{2},
\]
and combining this and Eq.~(\ref{eq:Torque-Im-Delta}) we find
\begin{equation}
N_{m}=-\frac{1}{2}\frac{kl_{21}^{2}|\tilde{\psi}(r_{w})|^{2}\Delta_{i}(ikv)}{(\Delta'-\Delta_{r}(ikv))^{2}+\Delta_{i}(ikv)^{2}}.\label{eq:Torque-GGJ}
\end{equation}
The denominators $|\Delta'-\Delta(ikv)|^{2}$ are minimized ($|\tilde{\psi}(r_{t})|^{2}$
is maximized) when the roots $\gamma$ corresponding to the spontaneous
mode, with $\Delta'=\Delta(\gamma_d)$, are closest to zero
in the complex plane. This occurs when $\pm\omega_{r}+kv$ is closest
to zero, i.e\@.~the frequency of the backward wave in the lab frame
$-\omega_{r}+kv$ is zero. See Fig.~\ref{fig:Locus-of-roots-RI}.   This allows maximum 
interaction with the zero frequency error field. Note that at this point $\Delta_i(ikv)$ is approximately zero so, by Eq.~(\ref{eq:Torque-GGJ}), $N_m$ is very close to zero where $v=\omega_r/k$.

\begin{figure}
\includegraphics[scale=0.4]{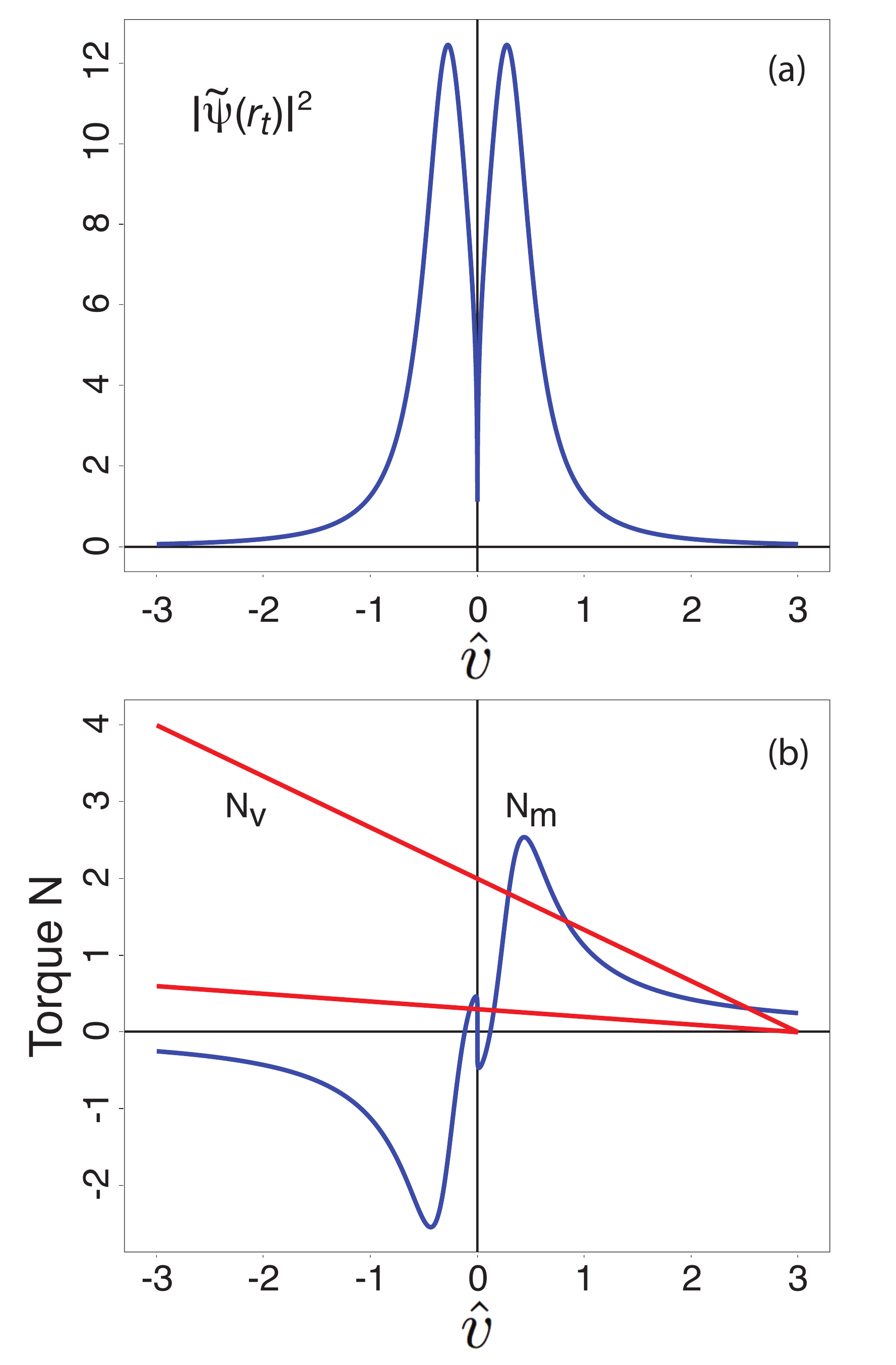}
\caption{Plot (a) of $|\tilde{\psi}(r_{t})|^{2}$ vs.~$\hat{v}\equiv kv\tau_{ri}$ for a weakly
stable RI mode with the Glasser effect, showing peaks at $v=\omega_r/k$, i.e.~$\hat{v}=\omega_{r}\tau_{ri}$. In (b) are shown the torque curve
$-N_{m}$ vs.~$\hat{v}$, with $N_{m}=0$ close to the phase velocity
$\hat{v}=\omega_{r}\tau_{ri}=0.12$; also shown are two lines $N_{v}=N_{0}(v_{0}-v)$
with different values of $N_{0}$. Note the differences between this figure and Fig.\ref{fig:TorqueCurveVR}.} 
\label{fig:TorqueCurveRI}
\end{figure}

\begin{figure}
\includegraphics[scale=1.0]{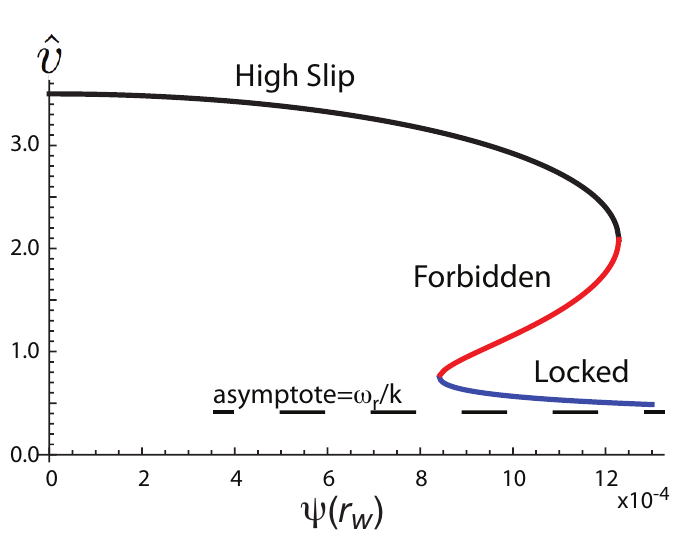}
\caption{Bifurcation diagram for the RI regime, indicating the locked (penetrated) state, the rapidly rotating state (high slip state) and the middle unstable state (labelled 'forbidden'.) Note the positive asymptote $v\rightarrow \omega_r/k$ as $\tilde{\psi}(r_w)$ increases.}
\label{fig:bifRI}
\end{figure}

The quantities $|\tilde{\psi}(r_{t})|^{2}$ and $N_{m}$ are shown
in Fig.~\ref{fig:TorqueCurveRI}, the latter with the viscous
torque $N_{v}(v)$.  Note that the reconnected flux is peaked away from $v=0$, approximately where $v=\omega_r/k$.  Note also that $N_{m}$ increases rapidly for $v\gtrsim\omega_{r}/k$,
so that the error fields tend to lock the plasma to just above the
phase velocity of the tearing mode $\omega_{r}/k$, as shown in Fig.~\ref{fig:TorqueCurveRI},
rather than to just above zero plasma velocity as in Fig.~\ref{fig:TorqueCurveVR}.  This effect produces a positive asymptotic value of plasma rotation $v\rightarrow \omega_r/k$ for the locked state as boundary field is increased, as shown in Fig. \ref{fig:bifRI}. The upper line shows three equilibria,
but with the leftmost intersection having $v\gtrsim\omega_r/k$
rather than $v\gtrsim 0$. For spontaneous modes which are
very weakly damped and have appreciable phase velocities, the locked
state can have significant plasma rotation. Also note that the intersection
with the lower line also indicates three equilibria, but with two
{\em negative} values of $v$, and one with $v\gtrsim\omega_r/k$.
All of these locked states with finite velocity exist in spite of the fact that the magnetic field driven by the
error field is steady, i.e.~locked to the error field. This effect is related to the fact that for RI (but not VR) resistive
wall tearing modes, small rotation can destabilize the modes by maximizing
the coupling of the tearing modes to the resistive wall\cite{FinnGerwinModeCoupling},
even for $D_s=0$.

We have formulated the inner layer equations in the RI regime with
$p'$, curvature and parallel dynamics, but ignoring perpendicular
compression and perpendicular particle transport, in order to determine
the importance of the last two effects on the stability and propagation of the spontaneous
modes. Using a scaling specific to this constant-$\psi$ regime, we
obtain
\begin{equation}
\frac{d^{2}W}{d\xi^{2}}-\xi^{2}W+\frac{GQ^{2}}{Q^{2}+b^{2}\xi^{2}}W=-\left(1+\frac{Gb^{2}}{Q^{2}+b^{2}\xi^{2}}\right)\xi,\label{eq:RI-eq-for-W}
\end{equation}
where $\tilde{\phi}=-i\alpha\delta^{3}\tilde{\psi}_{0}W(\xi)/\rho_{0}\eta$,
$r-r_{t}=\delta\xi$, $\delta=(\rho_{0}\gamma\eta)^{1/4}/(k_{||}'B)^{1/2}$, $G=-2m^{2}B_{\theta}^{2}(r_{t})p'(r_{t})\delta^{2}/B_{0}^{2}r_{t}^{3}\rho\gamma^{2}\propto D_{s}$,
and $Q=\gamma\tau_{ri}$. Also, we have $G=G_{0}/Q^{3/2}$ and $b^{2}=b_{0}^{2}Q^{1/2}$.
The effect of toroidicity is included by taking $G_{0}\rightarrow(1-q(r_{t})^{2})G_{0}$ as above,
leading to $G_{0}<0$ for $p'(r_{t})<0$ and $q(r_{t})>1$. The constant-$\psi$
small parameter is $\epsilon=\delta\Delta'\sim\eta^{2/5}$. We assume
$p'\sim\epsilon$, leading to $G\sim b\sim1$. The matching to the
outer region gives
\begin{equation}
\Delta'=\frac{\delta\gamma}{\eta}\int_{-\infty}^{\infty}\left(1-\xi W\right)d\xi=\frac{\delta\gamma}{\eta}\Delta_{s}(Q)=\gamma^{5/4}\tau_{ri}^{5/4}\Delta_{s}(Q)=Q^{5/4}\Delta_{s}(Q)\doteq\Delta(Q).\label{eq:Matching}
\end{equation}
The factors $1/(Q^{2}+b^{2}\xi^{2})\propto1/(\gamma^{2}+k_{||}^{2}c_{s}^{2})$
in Eq.~(\ref{eq:RI-eq-for-W}) represent sound wave propagation parallel
to $\mathbf{B}$; the factor $Q^{2}/(Q^{2}+b^{2}\xi^{2})$ on the
left is stabilizing; the term $\propto G$ on the right is destabilizing
but diminished by the $1/(Q^{2}+b^{2}\xi^{2})$ factor. 

Equation (\ref{eq:RI-eq-for-W}) and therefore $\Delta_{s}(Q)$ are
invariant under the symmetry $G_{0}\rightarrow\lambda G_{0},\,\, b_{0}\rightarrow\lambda^{1/2}b_{0},\,\, Q\rightarrow\lambda^{2/3}Q$.
This symmetry has two invariant quantities $G_{0}/b_{0}^{2}\propto p'/\Gamma p$
and $G=G_{0}/Q^{3/2}$. For $G_{0}$ positive (unfavorable curvature)
and sufficiently large, the $\Delta_{s}(Q)$ curve on the real $Q$
axis has poles (not shown) corresponding to localized unstable electrostatic
resistive interchanges $Q_{n}$. These modes are stabilized ($Q_{n}\rightarrow0$)
for sufficiently large sound speed ($G_{0}/b_{0}^{2}$ small enough.)
Past this threshold we observe numerically the relation $\Delta_{s}(Q)\approx A-B/Q^{3/2}$,
which shows 
\[
\Delta_{s}(Q)=\Delta_{s}(Q=\infty)-\frac{G_{0}}{Q^{3/2}}K\left(\frac{G_{0}}{b_{0}^{2}}\right)
\]
for some function $K$. A fit to the numerical data for large sound
speed $G_{0}/b_{0}^{2}\ll1$ gives
\begin{equation}
\Delta_{s}(Q)=2.12-\frac{2.77G_{0}}{Q^{3/2}}\left(1+0.65\frac{G_{0}}{b_{0}^{2}}\right).\label{eq:RI-final-result}
\end{equation}
This with Eq.~(\ref{eq:Matching}) shows the form of Refs.~\cite{GGJ1,GGJ2},
namely $\Delta(Q)=AQ^{5/4}-BQ^{-1/4}$, where $B\propto G_{0}\propto D_s$.
The unfavorable curvature ($G_{0}>0$) results of Ref.~\cite{FinnManheimer}
are recovered. In particular, these results show that $Q\sim(G_{0}/|\Delta'|)^{4}$
as $\Delta'\rightarrow-\infty$. 

The fit in Eq.~(\ref{eq:RI-final-result}) works well in the favorable
curvature case ($G_{0}<0$) also, again for sufficiently large sound
speed (small $|G_{0}/b_{0}^{2}|$.) Unlike in the unfavorable curvature
case, there is no evidence of unstable electrostatic resistive interchanges.
That is, these modes have complex $Q$ and do not show up as poles
in $\Delta(Q)$ on or near the real axis. These results for favorable
curvature show that the behavior observed in Refs.~\cite{GGJ1,GGJ2}
holds qualitatively using only parallel dynamics, without including
the divergence of the $E\times B$ drift and perpendicular particle
transport. In particular, there are complex roots for $\Delta'<\Delta_{min}=\min(\Delta(Q))\sim|G_{0}|^{5/6}$
and these roots become stable for $\Delta'<\Delta_{crit}\sim|G_{0}|^{5/6}$.
The locus of roots is shown in Fig.~\ref{fig:Locus-of-roots-RI}. 

The quantities $|\tilde{\psi}(r_{t})|^{2}$, $-N_{m}$ and $N_{d}$
as functions of the plasma rotation rate $v$ shown in Fig.~\ref{fig:TorqueCurveRI}
for $\Delta'\lesssim\Delta_{c}$ can be compared with those in Fig.~\ref{fig:TorqueCurveVR}.
For such values of $\Delta'$ the mode is weakly stable and the torque
is peaked just to the right of $v=\omega_{r}/k$, and \emph{negative}
to the left of this rotation value. The slope of $N_{m}(v)$ is steep
near $v=\omega_{r}/k$, so the locked state is just to the right of
the phase velocity. As in the VR case in Fig.~\ref{fig:TorqueCurveVR}
(with $\omega_{r}=0$), there can be two other roots, the rapidly rotating
state and an unstable middle root.

\begin{figure}
\includegraphics[scale=0.4]{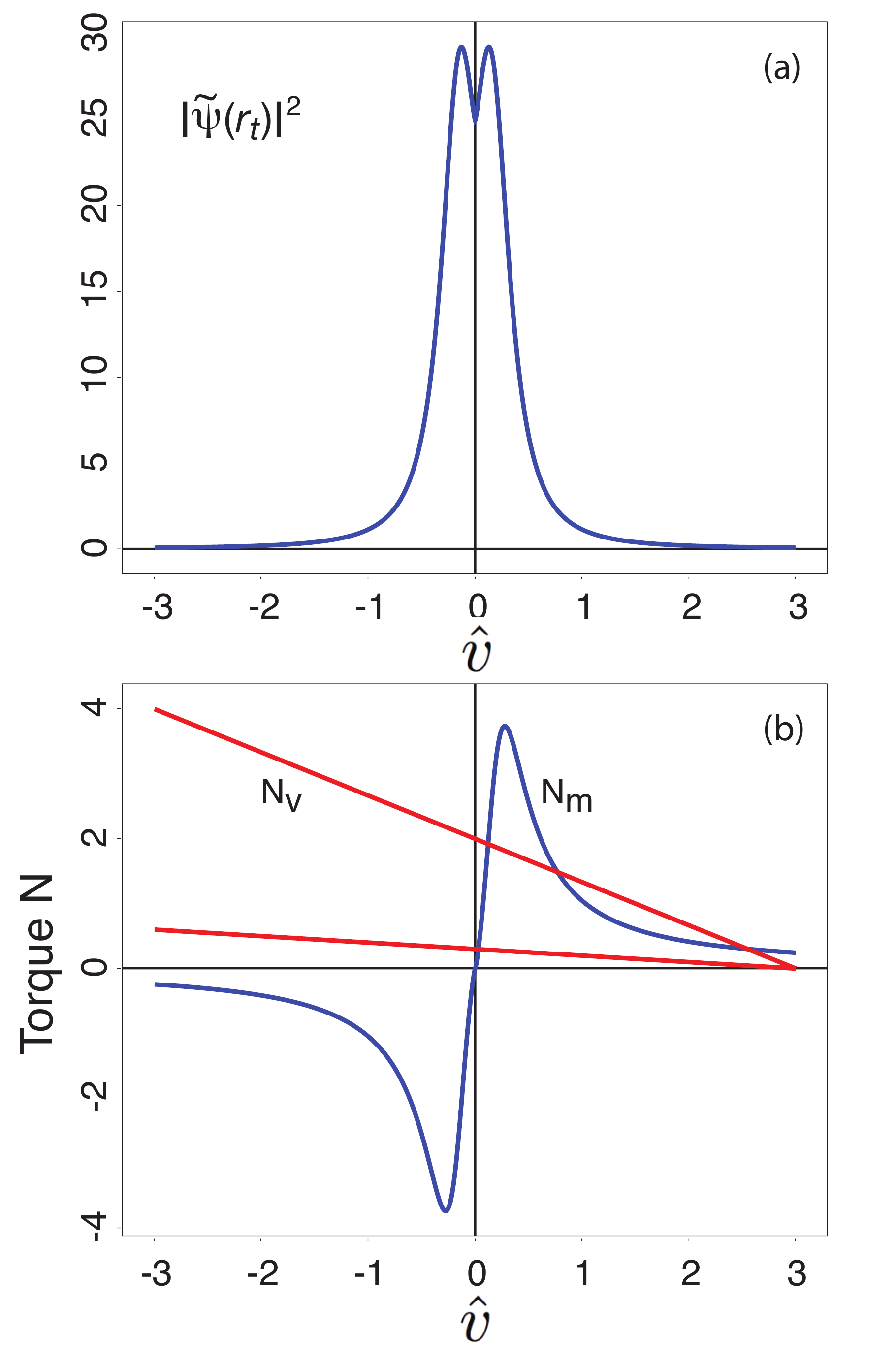}
\caption{The reconnected flux magnitude (a) $|\tilde{\psi}(r_{t})|^{2}$ in
the RI regime with $G_{0}=0$ shows peaks at finite $\hat{v}=kv\tau_{ri}$,
corresponding to complex roots, but the torque $N_{m}$ shown in (b)
goes to zero only at $v=0$, so locking to a nonzero velocity $v$
does not occur.}
\label{fig:TorqueCurveRI2}
\end{figure}

It is instructive to consider the case with $p'(r_{t})=0$ ($G_{0}=0$)
in the RI regime. As shown in Fig.~\ref{fig:TorqueCurveRI2},
the reconnected flux $|\tilde{\psi}(r_{t})|^{2}$ in the RI regime
with and $\Delta'\lesssim0$ also has two peaks at $kv\tau_{ri}\neq0$,
corresponding to stable complex conjugate roots with $Q\sim|\Delta'|^{4/5}e^{\pm4\pi i/5}$.
However, the torque curve is not qualitatively different from the
VR curve in Fig.~\ref{fig:TorqueCurveVR} for zero pressure
gradient, since $N_{m}\sim-\text{Im}(ikv)^{5/4}=-\text{sgn}(kv)|kv|^{5/4}\sin(5\pi/8)$,
which goes to zero only at $v=0$. It can easily be seen that the simple
form for the VR regime (without pressure gradient) $N_{m}\sim-v/(v_{1}^{2}+v^{2})$(
as in Refs.~\cite{Fitzpatrick1,Fitzpatrick2,BoozerRFA,FitzpatrickEtAl93})
does not hold quantitatively.

\section{Viscoresistive (VR) regime }

For the VR regime, with pressure gradient, curvature and parallel
dynamics, but without perpendicular compression or particle transport,
the inner region equation for the streamfunction $W$ takes the form

\begin{equation}
\frac{d^{4}W}{d\xi^{4}}+\xi^{2}W-\frac{GQ^{2}}{Q^{2}+b^{2}\xi^{2}}W=\left(1+\frac{Gb^{2}}{Q^{2}+b^{2}\xi^{2}}\right)\xi,\label{eq:W-equation-in-VR}
\end{equation}
where $r-r_{t}=\delta\xi$ with $\delta=(\eta\mu/\alpha^{2})^{1/6}$,
$Q=\gamma\tau_{vr}$, $G\propto D_{s}$ is defined as in the RI case,
and $b=\alpha\delta c_{s}\tau_{vr}/B$. The matching condition is
\begin{equation}
\Delta'=\frac{\delta\gamma}{\eta}\int_{-\infty}^{\infty}\left(1-\xi W\right)d\xi=\frac{\delta\gamma}{\eta}\Delta_{s}(Q)=\gamma\tau_{vr}\Delta_{s}(Q)=Q\Delta_{s}(Q)\doteq\Delta(Q).\label{eq:Matching-in-VR}
\end{equation}
We also have 
\[
G=\frac{G_{0}}{Q}.
\]
and
\[
b=b_{0}.
\]
Equation (\ref{eq:W-equation-in-VR}) has the symmetry $G_{0}\rightarrow\lambda G_{0},\,\, b_{0}\rightarrow\lambda b_{0},\,\, Q\rightarrow\lambda Q$,
so the invariants are $G_{0}/b_{0}$ and $G_{0}/Q$. Notice that this
symmetry extends to $\lambda<0$, i.e.~$Q\rightarrow-Q,\,\, G_{0}\rightarrow-G_{0}$
(and trivially to $b_{0}\rightarrow-b_{0}$.) The quantity $\Delta_{s}$
is unchanged, but $\Delta(Q)\rightarrow-\Delta(Q)$. Thus the results
for unfavorable curvature ($G_{0}>0$) can be applied directly to
the favorable curvature ($G_{0}<0$) case.

\begin{figure}
\includegraphics[scale=0.3]{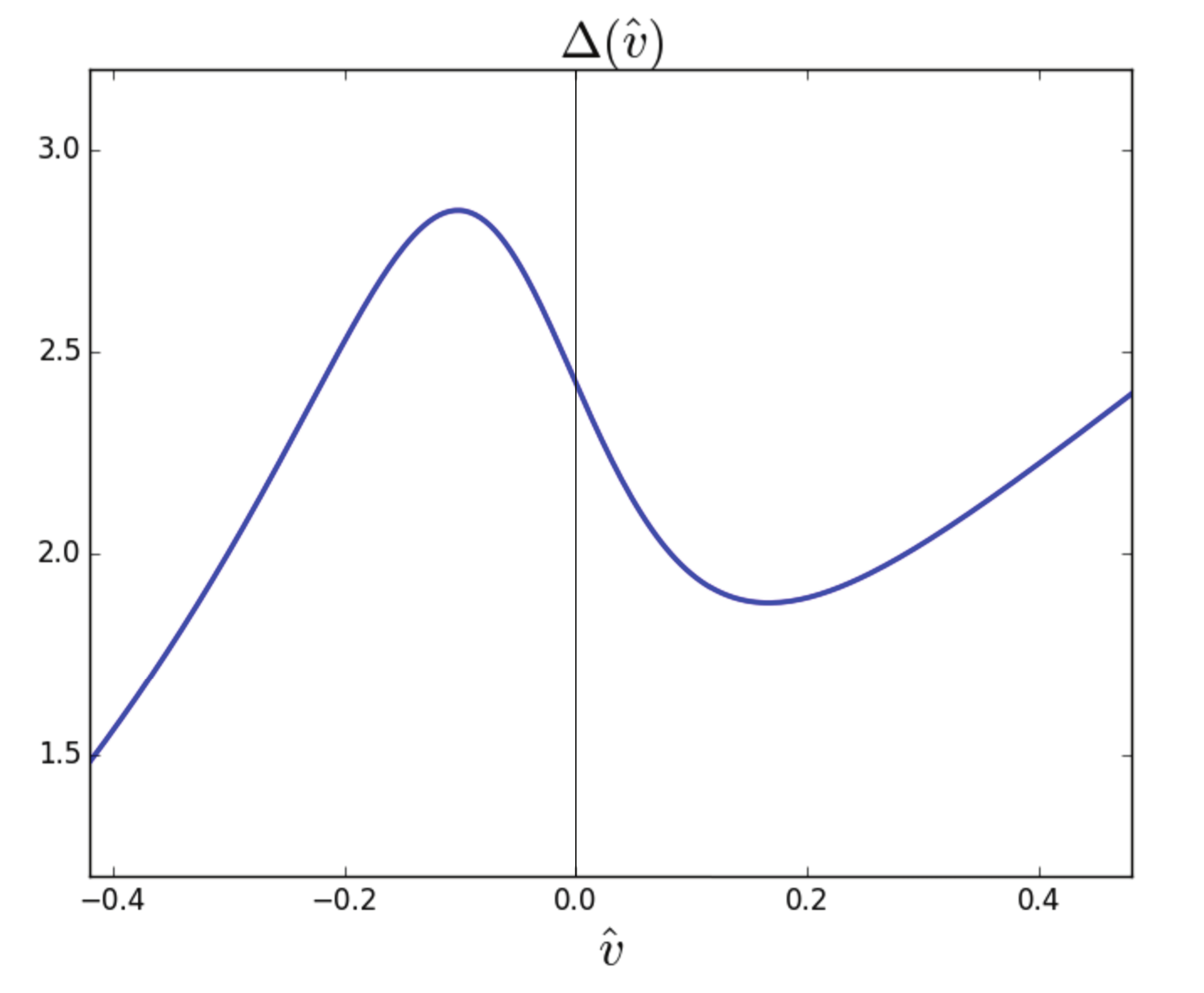}
\caption{Plot of $\Delta(Q)$ in the VR regime with favorable curvature for
$G_{0}/b_{0}=-1.6$.}
\label{fig:Deltas(Q)-for-VRGlasser}
\end{figure}

\begin{figure}
\includegraphics[scale=0.3]{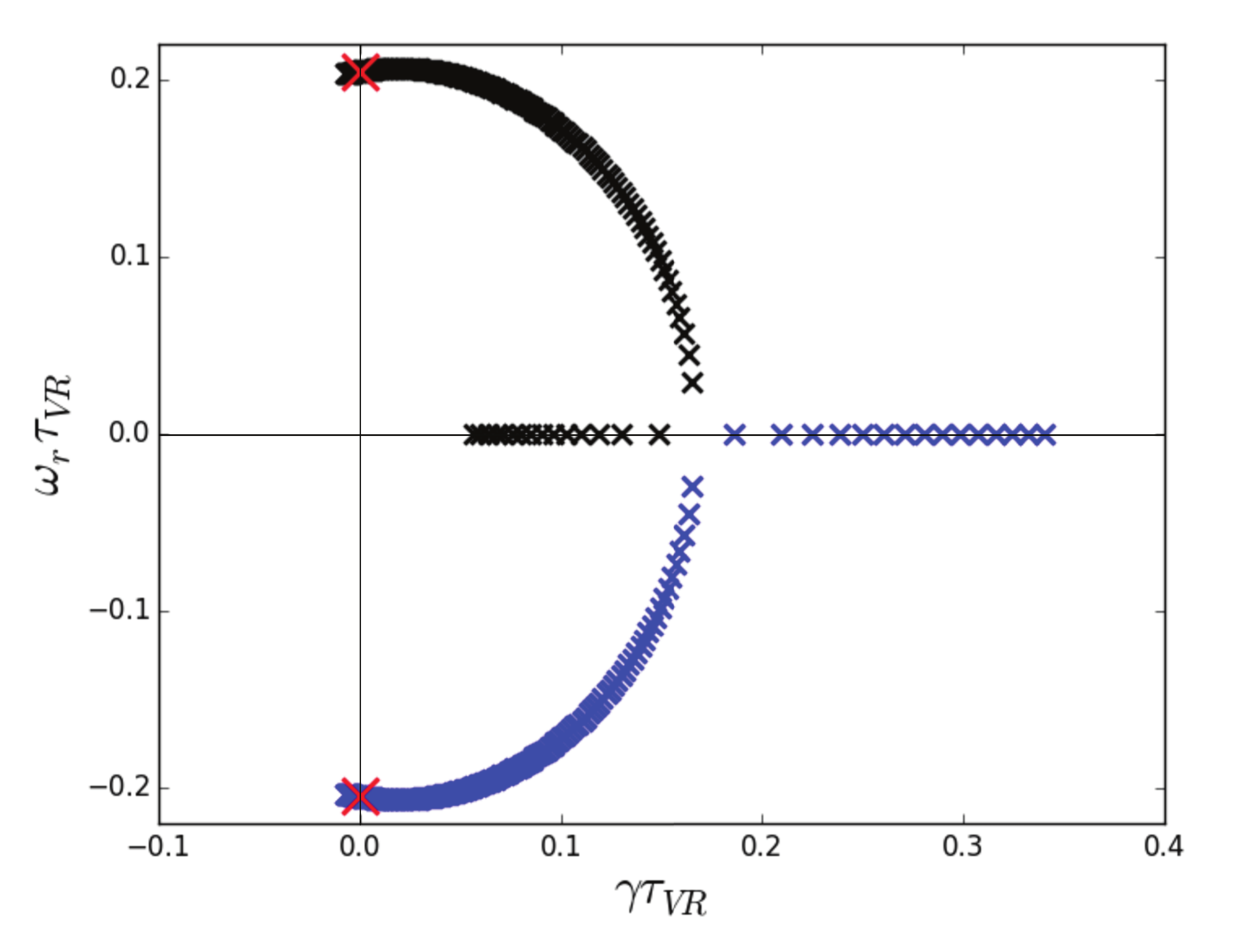}
\caption{Locus of roots showing that there is a Glasser effect in the VR regime.}
\label{fig:Locus-VR-Glasser}
\end{figure}

\begin{figure}
\includegraphics[scale=0.4]{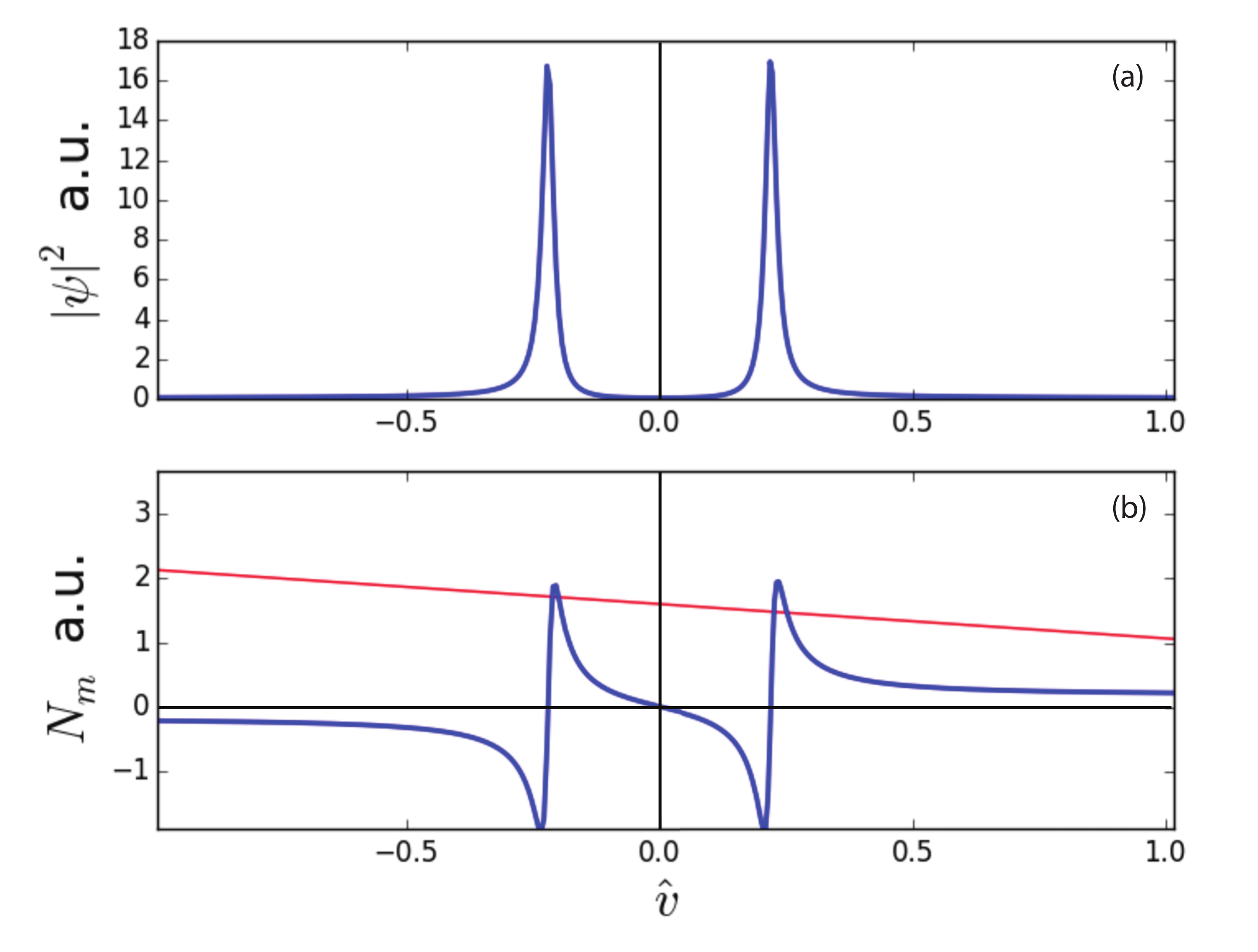}
\caption{Plot (a) of reconnected flux magnitude $|\tilde{\psi}(r_{t})|^{2}$
vs.~$\hat{v}=kv\tau_{vr}$ for a weakly stable VR mode with pressure gradient,
curvature, parallel dynamics, in the regime with a non-monotonic form
of $\Delta(Q)$.  Again, the reconnected flux magnitude is maximum near $v=\omega_r/k$.
In (b) are shown the torque curve $-N_{m}$ vs.~$\hat{v}$,
with $N_{m}=0$ at the phase velocity $\hat{v}=\omega_r\tau_{vr}=0.13$
with the two lines $N_{v}=N_{0}(v_{0}-v)$ having different values
of $N_{0}$. Here there can be as many as five equilibria, and the intersection
at the lower positive $\hat{v}$, the locked state, again has $\hat{v}\gtrsim\omega_r\tau_{vr}$ rather than
$\hat{v}\gtrsim0$.}
\label{fig:MajorTorqueCurveVR}
\end{figure}

We will describe the results first in terms of favorable curvature
($G_{0}<0)$. For large values of $|G_{0}/b_{0}|\sim|p'|/\sqrt{p}$
the plot of $\Delta_{s}(Q)$ on the positive real $Q=\hat{v}=\gamma\tau_{vr}$
axis shows poles, again corresponding to electrostatic resistive interchanges.
As $|G_{0}/b_{0}|$ decreases most of these modes become stabilized,
but the last two coalesce at $|G_{0}/b_{0}|=9.6$ for positive $Q$,
leaving a nonmonotonic $\Delta(Q)$ curve as shown in Fig.~\ref{fig:Deltas(Q)-for-VRGlasser}. 
This curve is nonmonotonic for the wide range $1.2<|G_{0}/b_{0}|<9.6$.
For $\Delta'$ just below the minimum on this curve, or just above
the maximum, there are complex roots. A plot of the locus of these
roots is shown in Fig.~\ref{fig:Locus-VR-Glasser}. We obtain a reconnected
flux curve and a torque curve shown in Fig.~\ref{fig:MajorTorqueCurveVR},
similar curves to those of in Fig.~\ref{fig:TorqueCurveRI},
for the RI regime. (The example in the VR regime in Fig.~\ref{fig:MajorTorqueCurveVR}
is for a much more weakly damped mode, leading to more peaked functions.) 

For unfavorable curvature ($G_{0}>0$), the symmetry $G_{0}\rightarrow-G_{0}$,
$Q\rightarrow-Q$ shows that similar complex roots occur, but they
are weakly growing when the roots for $G_{0}>0$ are weakly damped. 

\section{Conclusions}

Our major new result relates to the torque $N_{m}$ on the plasma
as a function of velocity $v$ for tearing mode regimes with real frequencies, such as the resistive-inertial (RI) regime with the Glasser effect. We find that $N_{m}(v)$, which is
concentrated in the tearing layer of the driven mode, is exactly zero
where $v=\omega_{r}/k$. This torque $N_{m}(v)$ is peaked just to
right of $v=\omega_{r}/k$, as shown in Figs.~\ref{fig:TorqueCurveRI} 
and \ref{fig:MajorTorqueCurveVR}. If we model the viscous torque
driving the plasma rotation as $N_{v}=N_{0}(v_{0}-v)$, then we find
that an error field can cause locking to a plasma velocity
$v$ with $v\gtrsim\omega_{r}/k$. Thus, the magnetic perturbation
locks to the error field at rest in the laboratory frame, but the
\emph{plasma} locks to a velocity just above the finite phase velocity $v\gtrsim\omega_{r}/k$ 
of the spontaneous tearing mode. Further, our results show that the plasma velocity asymptotes to this phase velocity as $\tilde{\psi}(r_w)\rightarrow \infty$ or $N_0 \rightarrow 0$.


We have also shown that in the VR regime with pressure gradient and
parallel dynamics in the tearing layers, there is a realistic and
wide range of parameters for which weakly stable tearing modes with
complex $\omega\approx\pm\omega_{r}+i\gamma$ can occur. That is,
there is a Glasser effect in the VR regime as well as in the RI regime\cite{GGJ1,GGJ2}.
In both regimes, the presence of real frequencies $\omega_{r}$ in
the plasma frame implies that the driven tearing mode has a very peaked
maximum amplitude where the plasma velocity $v$ equals the phase
velocity $\omega_{r}/k$.

Tearing modes in the presence of diamagnetic effects can also have
real frequencies $\omega_{r}\propto\omega_{*}$, with a reduction in
growth rate $\gamma$ \cite{Coppi-1,Coppi-2,Biskamp,FMA}. It is expected
that diamagnetic propagation will cause a similar locking of a plasma
to a rotating state, $v\gtrsim\omega_{r}/k$. For such modes, however,
there will be no $v\rightarrow-v$ symmetry because the modes do not
occur in complex conjugate pairs.

For sufficiently large error fields, the mode amplitude in the locked
state can be high enough to be in the Rutherford regime\cite{Rutherford},
where the magnetic island width $w$ is comparable to the tearing
layer thickness. In such cases, the linear theory for the mode (and
quasilinear theory for the torque) should still be qualitatively accurate.
For reasonable DIII-D parameters with moderate plasma rotation, we
find that the locking threshold is sufficiently small that the tearing
mode amplitude is below or comparable to that for the Rutherford regime,
so that the above conclusions are valid.

At even higher error field amplitudes, sound wave propagation can flatten the pressure around the island\cite{ScottEffect} and this
weakens the mode propagation. For DIII-D parameters with fast rotation,
the locking threshold is for a large error field, so that typically
the plasma will not lock. However, in the presence of such a large
error field, locking by a necessarily large error field can lead to
the formation of a quite large island, and the finite frequency effects
discussed in this paper, due to pressure gradient, may not be present. 

We have one caveat about the linear tearing mode regimes and the nonlinear regimes discussed above. These are useful for guidance but, for example, when analyzing modes in the VR regime, the mode may move into another regime as the velocity increases. Also, the correct regime may be different for the locked and unlocked states in the same bifurcation diagram.

These results provide a new viewpoint on the locking phenomenon: the locking force by itself leads to a plasma velocity equal to the phase velocity of the spontaneous mode. (The actual velocity of the plasma must be calculated using a balance the other torques, a momentum source due to neutral beams.) As an example of an application of this framework, our results suggest the possibility of applying error fields with
a spread in Fourier spectrum. Then locking may occur in each tearing layer to just above the phase velocity for its Fourier harmonic. If these phase velocities are not equal, then viscosity (e.g.~NTV) should lead to a sheared rotation between these mode rational surfaces. This possibility might be useful for maintaining
flow shear to help stabilize other modes.\\
\\
\\
\textbf{Acknowledgments.} The work of J.~M.~Finn was supported by
the DOE Office of Science, Fusion Energy Sciences and performed under
the auspices of the NNSA of the U.S.~DOE by LANL, operated by LANS
LLC under Contract No DEAC52-06NA25396. The work of D.P.~Brennan and A.J.~Cole was supported by the DOE Office of Science collaborative grants DE-SC0014005 and DE-SC0014119 respectively.

\end{document}